\documentclass[a4,12pt]{article}
\usepackage{amsfonts,amsmath,amssymb,epsf,cite}
\usepackage{graphicx,color}
\usepackage[dvipdfm,hypertex]{hyperref}

 \def\m{{\mu}}
 
 \def\n{{\nu}}

 \def\frac#1#2{{#1\over #2}}

 \def\s{\sqrt}

 \def\CO{{\cal O}}

 \def\CN{{\cal N}}
 \def\p{\partial}

 \def\de{\partial}

 \def\f {\frac}
 
 \def\ap{\alpha}

 \def\ddd{\cdot\cdot\cdot}
 \def\no{\nonumber \\}
 \def\la{\langle}
 \def\lb{\rangle}

 \def\m{{\mu}}
 
 \def\n{{\nu}}

 \def\frac#1#2{{#1\over #2}}

 \def\s{\sqrt}

 \def\CO{{\cal O}}

 \def\CN{{\cal N}}
 \def\p{\partial}

\def\be{\begin{equation}}
\def\ee{\end{equation}}
\def\ba{\begin{eqnarray}}
\def\ea{\end{eqnarray}}

\makeatletter
\renewcommand\section{\@startsection {section}{1}{\z@}%
                                   {-3.5ex \@plus -1ex \@minus -.2ex}%nn
                                   {2.3ex \@plus.2ex}%
                                   {\normalfont\large\bfseries}}
\renewcommand\subsection{\@startsection{subsection}{2}{\z@}%
                                     {-3.25ex\@plus -1ex \@minus -.2ex}%
                                     {1.5ex \@plus .2ex}%
                                    {\normalfont\bfseries}}

\begin{document}
\begin{titlepage}
\thispagestyle{empty}
\begin{flushright}
IPMU09-0134\\
KUNS-2239
\end{flushright}

\bigskip

\begin{center}
\noindent{ \large\textbf{Holographic Superconductor/Insulator Transition\\ at Zero Temperature}}\\
\vspace{2cm}
 \noindent{Tatsuma Nishioka$^{a,b}$\footnote{e-mail: nishioka@gauge.scphys.kyoto-u.ac.jp},
Shinsei Ryu$^{c}$\footnote{e-mail: sryu@berkeley.edu}
and Tadashi Takayanagi$^{b}$\footnote{e-mail: tadashi.takayanagi@ipmu.jp}}
\vspace{1cm}

{\it $^{a}$Department of Physics, Kyoto University, Kyoto, 606-8502, Japan\\
 $^{b}$Institute for the Physics and Mathematics of the Universe, \\
 University of Tokyo, Kashiwa, Chiba 277-8582, Japan\\
  $^{c}$Department of Physics, University of California, Berkeley, CA 94720, USA}
\end{center}

\vspace{0.5cm}
\begin{abstract}
We analyze the five-dimensional AdS gravity coupled to a gauge field and a charged scalar field.
Under a Scherk-Schwarz compactification, we show that the system undergoes
a superconductor/insulator
transition at zero temperature in $2+1$ dimensions
as we change the chemical potential.
By taking into account a confinement/deconfinement transition,
the phase diagram turns out to have a rich structure. We will observe
that it has a similarity
with the RVB (resonating valence bond) approach to high-$T_c$ superconductors via
an emergent gauge symmetry.
\end{abstract}

\end{titlepage}

%\newpage

%\begin{scriptsize}
%\tableofcontents
%\end{scriptsize}

%\newpage

\section{Introduction}
\label{intro} %\setcounter{equation}{0}
%\hspace{5mm}
The AdS/CFT correspondence \cite{Ma} is a very powerful method to analyze the phase structures
and phase transitions in
strongly coupled quantum many-body systems.
This is because these complicated
quantum problems are equivalently mapped to the study of their gravity duals, each of which is often
described by a classical Einstein gravity coupled to various matter fields.

A rich phase diagram can appear in quantum many-body systems
when there are many competing and mutually frustrating interactions.
For example,
one of the richest and most interesting phase structures has been known to appear
in strongly correlated materials, such as
the cuprate high-$T_c$ superconductors \cite{ReviewHTC} (see the Fig.\ \ref{fig:HTC}
in the main text),
and layered organic conductors.
In the phase diagram of high-$T_c$,
an insulator phase with the antiferromagnetic order
(called the Mott insulator) is located close to the superconductor phase.
This implies that the strong Coulomb repulsive interaction is behind
the pairing mechanism of the superconductivity:
It is highly likely that superconductivity is not driven by a conventional mechanism
due to simple attractive forces.
Instead the superconductivity emerges as
the best compromise among many competing ground states.
Adjacent to the Mott insulator and superconductor phases,
there appear the pseudo gap and strange metal phases,
where transport properties are highly unusual.
As we increase doping $x$ beyond the superconductivity phase,
there is the fermi-liquid phase.

The purpose of this paper is to study such a complex competition
of different quantum ground states in a simple setup of AdS/CFT.
Following the idea of the holographic superconductor \cite{HHHone,Gu,HKS,HRone,HHHtwo,GN,HaR,HeR},
we consider the Einstein gravity coupled to a $U(1)$ gauge field and a charged scalar field.
The $U(1)$ gauge field is interpreted as an R-charge and the scalar field is dual to an R-charged operator.
An important difference  from the earlier works
is that in our setup
the holographic superconductor is put in the five-dimensional AdS soliton background \cite{Wi,HM}.
As we compactify one of the space directions in this asymptotic AdS spacetime,
the AdS soliton is dual to a Scherk-Schwarz compactification of a four-dimensional conformal gauge theory,
such as the $\CN=4$ super Yang-Mills theory.
The dual field theory is thus (2+1)-dimensional as is so in the high-$T_c$ superconductors.

In the holographic superconductors,
there is a finite temperature superconductor-metal transition.
On the other hand,
the AdS soliton background decays into AdS black hole via a Hawking-Page transition \cite{HP},
which is dual to the confinement/deconfinement transition \cite{Wi}.
One of our main findings is that at zero temperature,
as we change the chemical potential,
there is an intervening phase between the holographic superconductor phase
and the confining phase,
which we call the AdS soliton superconductor phase
(refer to Fig.\ \ref{fig: schematic Phase in intro} in the main text).
While this phase is similar to the holographic superconductor phase
(labeled as the AdS BH superconductor phase in Fig.\ \ref{fig: schematic Phase in intro})
in the sense that it is the phase where the charged scalar condenses,
this phase has a much larger gap.
The confinement phase (labeled as the AdS soliton in Fig.\ \ref{fig: schematic Phase in intro})
is a phase with a mass gap and
an analogue of an insulator in electronic systems.
Our setup thus realizes, as we change the chemical potential,
an insulator-to-superconductor quantum phase transition.

We will compare these findings with the RVB (resonating valence bond)
approach to high-$T_c$ superconductors \cite{LNW}.
The RVB theory starts by taking into account
the effects of the strong Coulomb interaction
which is most operative near zero doping, $x\sim 0$.
The strong Coulomb repulsion forbids
any site to be occupied by more than one electron,
and thus serves effectively as a constraint on the physical Hilbert space.
This constraint can be written as a gauge constraint,
and therefore physics of high-$T_c$ superconductors near $x\sim 0$
can be written in the language of
an $SU(2)/U(1)/\mathbb{Z}_2$ gauge theory
(the gauge group depends on, at the level of the mean field theory,
the choice of a saddle point.)
The quasi-particles charged under the gauge field
is not an electron, but an emergent entity
which carries a fraction of the quantum number of electrons.
The abnormal phases in high-$T_c$,
i.e., the phases which are not well-described in terms of electrons,
can then be potentially described
in terms of these emergent particles when they are deconfined,
whereas confinement thereof gives rise to
conventional phases.
The high-$T_c$ superconductor thus provides a venue where
there is a complex interplay between confinement/deconfinement
physics and superconductivity
in (2+1) dimensions.

This paper is organized as follows:
in section \ref{BubbleSC},
we will show that an superconductor/insulator phase transition
will occur in the AdS soliton background.
In section \ref{AdS5 Charged Black Hole and Superconductor}, 
we will analyze the superconductor phase in
the AdS$_5$ black hole.
In section \ref{Phase Structure}, 
we will consider the phase diagram of our system and discuss its
relation to the RVB approach to the high-$T_c$ superconductors.
In section \ref{Conclusions}
we will summarize conclusions.

\section{Holographic Superconductor/Insulator Transition at Zero Temperature}
\label{BubbleSC}
%\hspace{5mm}
Let us begin with the five-dimensional Einstein-Maxwell-scalar theory:
\begin{align}
 S&= \int d^5x \s{-g} \left(R + \f{12}{L^2} - \f{1}{4}F^{\m\n}F_{\m\n}
 - |\nabla_\m \Psi - iqA_\m \Psi|^2 - m^2|\Psi|^2 \right) \ . \label{EMSaction}
\end{align}
Notice that almost the same system appears in the AdS$_5\times$S$^5$ compactification
dual to $\CN=4$ super Yang-Mills\footnote{Though in many string theory setups
there exist Chern-Simons terms, we will not consider it in this paper as they do not
change our solutions. Recently a new possibility of instability in the presence of
Chern-Simons terms has been found in \cite{Ooguri}.}. There, the $A_\mu$ and $\Psi$ can be dual to a R-charge current
and a R-charged operator with charge $q$, though we will not further
pursuit a string theory interpretation in this paper. In section 4.3, we will show that we can
consistently embed our phase transition into the $\CN=1$ superconformal field theory dual to
AdS$_5\times T^{1,1}$.

In this theory, our setup is defined by an asymptotically AdS spacetime which approaches to $R^{1,2}\times S^1$ near the
boundary. In this section we only consider the case of zero temperature.
We impose the anti-periodic boundary condition for fermions (i.e. Scherk-Schwarz boundary condition)
in the $S^1$ direction. When the gauge field is vanishing,
the most stable configuration which satisfies this property is known as the AdS soliton \cite{Wi,HM}
given by the metric
\begin{align}\label{bubblesol}
 ds^2&= L^2\f{dr^2}{f(r)} + r^2( -dt^2 + dx^2 + dy^2 ) +
 f(r)d\chi^2 \ ,\\
  f(r) &= r^2 - \f{r_0^4}{r^2} \ .  \nonumber
\end{align}
This can be obtained by double Wick rotating
the AdS Schwarzschild black hole.
In order to have a smooth geometry we need to impose the periodicity
$\chi\sim\chi+\f{\pi L}{r_0}$ for the Scherk-Schwarz circle. If we extract the coordinates
$(r,\chi)$, the geometry looks like a cigar, whose tip is given by $r=r_0$.

Now we take into account the coupling of this system to the gauge field.
We can easily find a simple solution with the constant gauge potential
$A_t=\mu$.\footnote{In general, it is possible that we have other solutions which look more non-trivial.
However, we do not discuss it because in this paper we are always working
within the approximation where we can ignore the backreaction of
gauge field and scalar to the metric.}
 Notice that as opposed to the AdS black holes which require $A_t=0$ at the
horizon, the boundary condition is chosen
so that the gauge field is non-singular at the tip.

\subsection{Superconductor/Insulator in AdS$_5$ Soliton}

In the presence of charged scalar field as in (\ref{EMSaction}), the phase structure can be more non-trivial as
the scalar field can condense. We would like to analyze this below assuming that the backreaction of
the gauge field and scalar to the metric (\ref{bubblesol})
is negligible. We focus on the solutions depending on only radial coordinate as
follows: \begin{align}
 A&= \Phi(r) dt \ , \qquad \Psi = \Psi(r) \ .
\end{align}
Moreover, we take $\Psi$ to be real due to the Maxwell
equation. We can also set $L=1$ and $r_0=1$ without losing generality.
At zero temperature the system depends on the parameters $(q,\mu,m^2)$.

The equations of motion are
\begin{align}
 &\Psi'' + \left( \f{f'}{f} + \f{3}{r}\right)\Psi' + \left( -\f{m^2}{f}
 + \f{q^2\Phi^2}{r^2f}\right)\Psi = 0 \ , \\
 & \Phi'' + \left( \f{f'}{f} + \f{1}{r} \right)\Phi' -
 \f{2q^2\Psi^2}{f}\Phi = 0 \ .\label{HCEOM}
\end{align}

We have to impose the boundary condition at the tip $r=r_0$ and the boundary
$r=\infty$ to solve the above equations. We will concentrate on the
case where $m^2$ is given by $m^2=-\f{15}{4}$ which satisfies the BF bound $m^2>-4$.
Though this is just for simplicity, we can still extend our analysis to general
values of $m^2$.
Near the boundary, the solutions behave as
\begin{align}
 \Psi& = \f{\Psi^{(1)}}{r^{3/2}} + \f{\Psi^{(2)}}{r^{5/2}} + \dots  \
 ,\\
 \Phi&= \mu - \f{\rho}{r^2} + \dots \ . \label{gaugeex}
\end{align}
At this value of $m^2$, there are two alternative
descriptions since
both terms,  proportional to
$\Psi^{(1)}$ and $\Psi^{(2)}$, respectively,
become normalizable. We defined the corresponding dual operators
by $\CO_1$ and $\CO_2$ whose conformal dimensions are give by $\Delta=3/2$ and $\Delta=5/2$, respectively.

On the other hand, at the tip these behave as
\begin{align}
 \Psi&= a + b\log (r-r_0) + c(r-r_0) + \dots \ , \no
 \Phi&= A + B\log (r-r_0) + C(r-r_0) + \dots \ .   \label{neumann}
\end{align}
Therefore, we impose the Neumann-like boundary condition $b=B=0$ to take every physical
quantities finite.

In (\ref{gaugeex}), the constants $\mu$ and $\rho$ are holographically dual to the chemical potential
(gauge potential) and the charge density, respectively.

As is clear from (\ref{HCEOM}), the equations of motion have the scaling symmetry
$(\Phi,\Psi,\mu,q)\to
(\lambda\Phi,\lambda\Psi,\lambda\mu,q/\lambda)$. By taking $\lambda\ll
1$ (the probe limit),
we can indeed neglect the back reactions of $\Phi$ and $\Psi$ to the metric as in the standard
holographic superconductors \cite{HHHone,HHHtwo}. In other words, this probe approximation
is justified when $\mu\ll 1$ and $q\gg 1$ with $\mu q$ kept finite.
Once we take this probe limit, the functional form of physical quantities
up to the overall scaling only depend on the combination $\mu q$ in this probe limit.
Thus below we will simply set $q=1$.

\begin{figure}[htbp]
 \begin{center}
  \includegraphics[height=3.5cm]{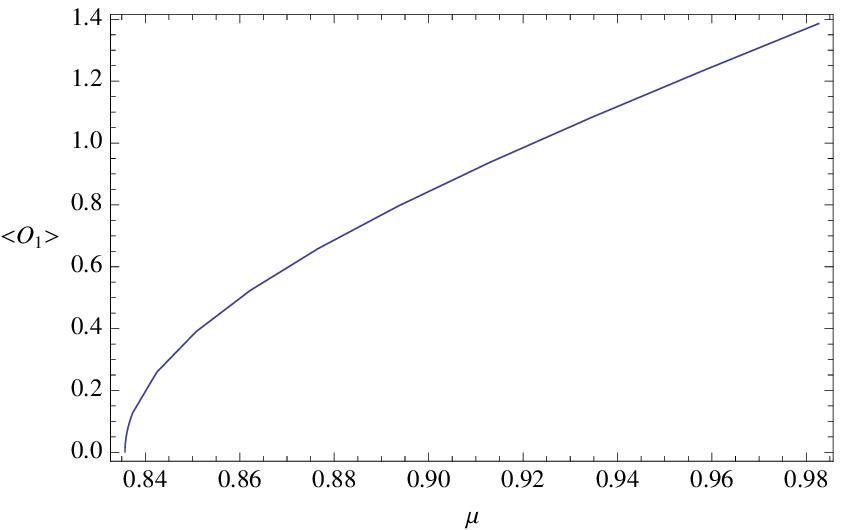}
  \hspace{1cm}
  \includegraphics[height=3.5cm]{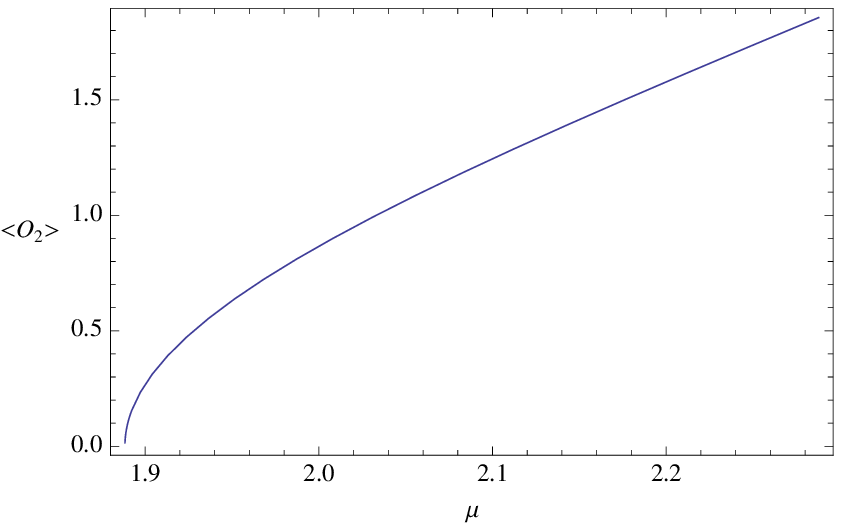}
 \end{center}
 \caption{The condensations of the scalar operators $\langle \CO_1 \rangle$ (left) and $\langle \CO_2 \rangle$ (right).
 \label{fig:Condensate} }
\end{figure}

In the AdS/CFT dictionary, the asymptotic behavior of $\Phi$ gives the
chemical potential $\m$ and charge density $\rho$ in the dual field
theory. The scalar operator $\CO$ coupled to the scalar field $\Psi$
is (up to a normalization)
\begin{align}
 \langle \CO_i \rangle &= \Psi^{(i)} \ , \quad i = 1,2
\end{align}
with the boundary condition $\epsilon_{ij}\Psi^{(j)}=0$.
We can plot $\langle \CO_i\rangle$ with respect to the chemical potential
$\m$ by numerical calculations and find that the condensation occurs if $\mu>\mu_{i}$
($\mu_1= 0.84$ and $\mu_2=1.88$) as in Fig.\ \ref{fig:Condensate}.
Notice that for any values of $q$, the probe approximation offers us exact
critical values $\mu_{1,2}$ and behaviors near the the critical points, where the
back reactions are highly suppressed.

\begin{figure}[htbp]
 \begin{center}
  \includegraphics[height=3.5cm]{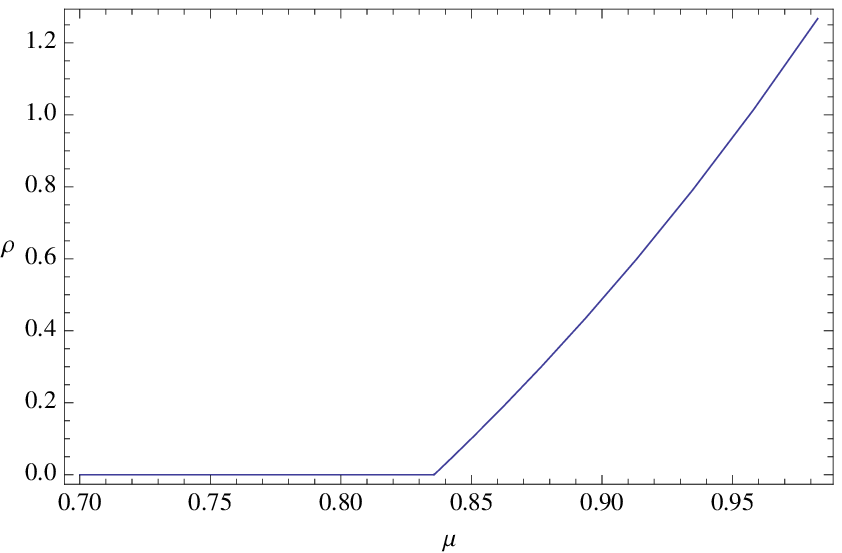}
    \hspace{1cm}
  \includegraphics[height=3.5cm]{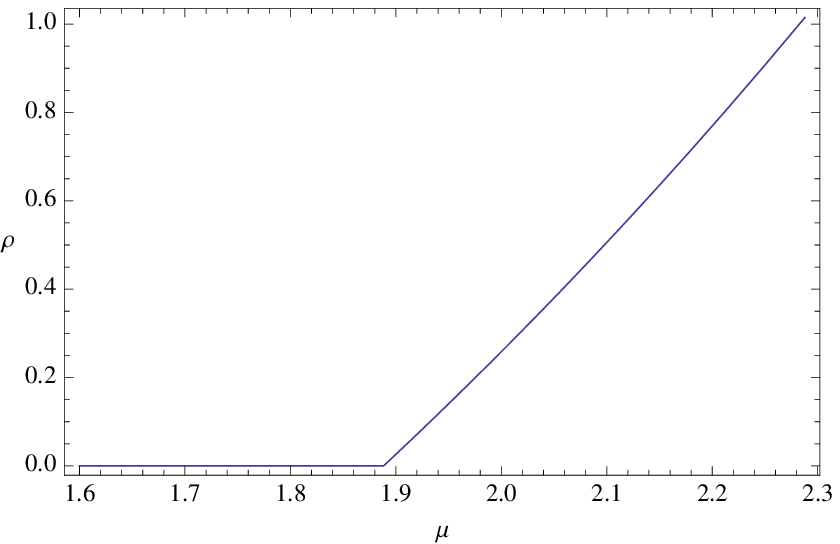}
 \end{center}
 \caption{The charge density $\rho$ plotted as a function of $\mu$
 when $\la \CO_1\lb\neq 0$ (left) and
$\la \CO_2\lb\neq 0$ (right).
Its derivative jumps at the
 phase transition point.
 \label{fig:rhomu}}
\end{figure}

In this way, we find a phase transition triggered by
the scalar field condensation in the AdS soliton background.
We also find that it is a second order phase transition as is clear if
we plot the charge density $\rho\ (=\de{\Omega}/\de{\mu})$ as
a function of $\mu$ as shown in Fig.\ \ref{fig:rhomu}.
When $\mu$ is small, the system is described by the AdS soliton solution itself. This is interpreted as the
insulator phase as this system has a mass gap, which is due to the confinement in the (2+1)-dimensional gauge
theory viewpoint via the Scherk-Schwarz compactification. On the other hand, for a larger $\mu$, it undergoes a
phase transition and is expected to reach a superconductor (or superfluid) phase. Thus we can regard this as the
holographic realization of superconductor/insulator transition.

The mechanism of this transition is similar to the standard
holographic superconductors \cite{HHHone,HHHtwo}, but is different in that our
phase transition occurs even at strictly zero temperature. We also give an explanation of
the transition by relating it to the Schr{\" o}dinger problem in the appendix A.

If we impose the periodic boundary condition instead of the anti-periodic
one for fermions, the story will change largely because the AdS soliton phase is not available
and the confinement will not occur at zero temperature. In this case, we need to consider
a zero temperature limit of holographic superconductor in AdS black holes, which has been
analyzed in \cite{GQ,GR,ZT,Ko}. We will not discuss this case in this paper.

 So far we only consider the zero temperature system.
At finite temperature $T$, it is described by compactifying the Euclidean
time direction. When $T$ is greater than a certain value $T_c$, a phase transition like
confinement/deconfinement transition occurs as we will see later. Thus our argument of
phase transition can equally be applied to the finite temperature case
as long as we have $T<T_c$ and we can ignore the back reaction.

\subsection{Conductivity}

We can holographically calculate the conductivity $\sigma(\omega)$ by solving the equation of motion of
$A_x\propto e^{-i\omega t}$:
\begin{align}
 &A_x'' + \left( \f{f'}{f} + \f{1}{r} \right)A_x' + \left(
 \f{\omega^2}{r^2 f} - \f{2q^2\Psi^2}{f}\right) A_x = 0 \ , \label{axeom}
\end{align}
 by requiring the Neumann boundary condition at the tip $r=r_0$ as in (\ref{neumann}).
 Looking at the asymptotic behavior near the boundary $r\to \infty$
\be
A_x=A_x^{(0)}+\f{A_x^{(1)}}{r^{2}}+\f{A_x^{(0)}\omega^2}{2}\f{\log\Lambda r}{r^2}+\ddd\ ,
\ee
the holographic conductivity is found as follows \cite{HRone}
\be
\sigma(\omega)=\f{-2iA_x^{(1)}}{\omega A_x^{(0)}} + \f{i\omega}{2} \ .
\ee

\begin{figure}[htbp]
 \begin{center}
 \includegraphics[height=4cm]{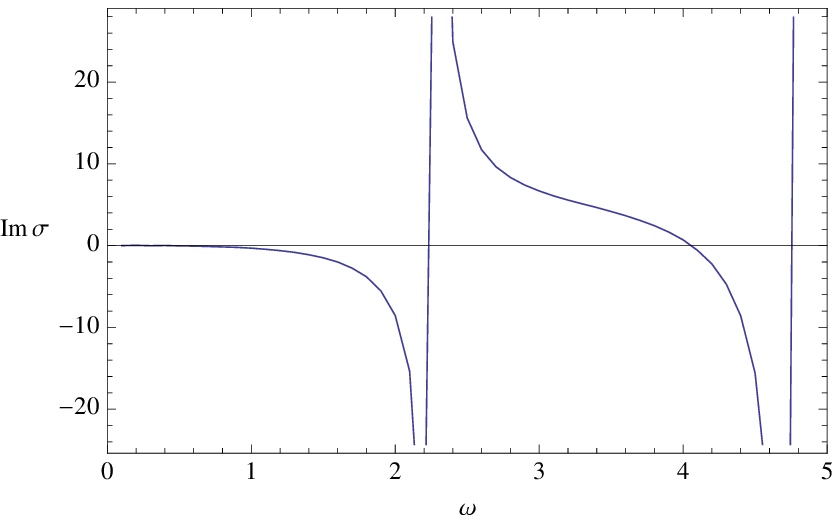}
  \hspace{1cm}
  \includegraphics[height=4cm]{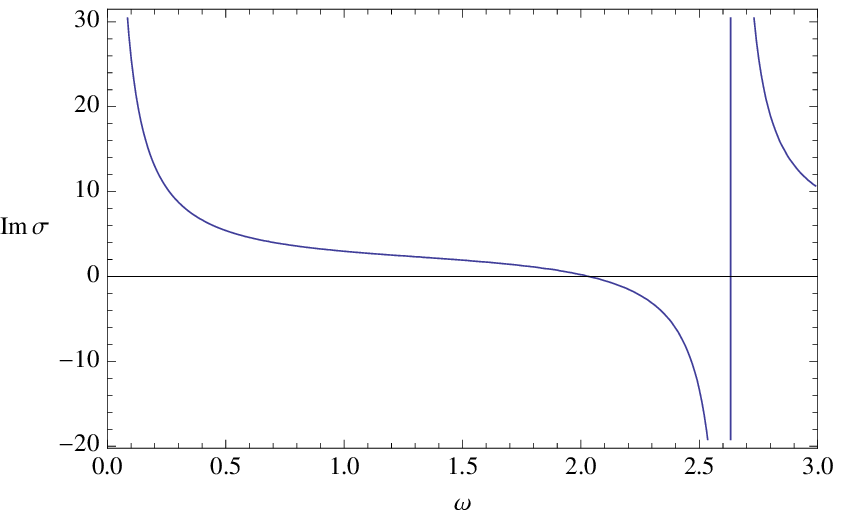}
 \end{center}
 \caption{The imaginary part of the conductivity for the AdS soliton without a scalar
condensation
$\la \CO_{1,2}\lb=0$
 (left) and with a scalar
condensation
$\la \CO_1\lb\neq 0$.  We employed the background
with $\rho=0.0094$ and $\mu=0.84$ in the right graph.
 \label{fig:BubbleSC} }
\end{figure}

First let us consider the AdS soliton background (\ref{bubblesol}) without the scalar condensation.
Since there is no horizon, the equation of motion with the boundary condition at the tip
requires that $\f{A_x^{(1)}}{A_x^{(0)}}$ is real. Therefore the real part of the conductivity vanishes.
This means that there is no dissipation and is consistent with the absence of
horizon. The imaginary part of $\sigma(\omega)$ is plotted in the left graph of Fig.\ \ref{fig:BubbleSC}.
In this way, we can identify the AdS soliton with an insulator.
We also notice there are poles periodically at the points where $A_x^{(0)}$ vanishes. These correspond to
normalized modes dual to vector operators. In the paper \cite{FLRT}, this setup has been
used to realize the fractional quantum Hall effect by adding D7-branes wrapped on $S^5$,
which is a basic example of topological insulators (see \cite{AWT} for its back reacted
supergravity solution).
On the other hand, if we calculate $\sigma(\omega)$ for the AdS soliton with the scalar condensation by taking
$\mu>\mu_{1,2}$, then the imaginary part of  $\sigma(\omega)$ is obtained as
in the right graph of Fig.\ \ref{fig:BubbleSC}.
When $\omega$ is large, the behavior looks very similar to the one for the AdS soliton
without the scalar condensation.
However, in the present case, we can observe a pole even at $\omega=0$. The Kramers-Kronig relation
\be
\mbox{Im}\,\sigma(\omega)=\f{1}{\pi}P\int^\infty_{-\infty}
d\omega' \f{\mbox{Re}\,\sigma(\omega')}{\omega-\omega'}\ .   \label{KrKn}
\ee
argues
that Re\,$\sigma(\omega)$ has a delta functional support. Thus this should be identified with the
superconductivity. Defining the superfluid density $n_s$ by $\mbox{Im}\, \sigma(\omega)\sim \f{n_s}{\omega}$ in the
$\omega\to 0$ limit, we obtain the behavior $n_s\sim
20\mu_1(\mu-\mu_1)$
and  $n_s\sim 1.2 \mu_2 (\mu-\mu_2)$ for
each case near the transition.

\section{AdS$_5$ Charged Black Hole and Superconductor}
\label{AdS5 Charged Black Hole and Superconductor}
At high temperature, we expect that the AdS soliton
background will decay into the AdS black hole
via a Hawking-Page transition, which is dual to
the confinement/deconfinement transition \cite{Wi}.
It is well expected that this should be true even in the presence of scalar condensations.
Therefore, here we want to
analyze the properties of five-dimensional charged AdS black hole solutions and their
superconductor phase transition. We start with the Einstein-Maxwell
theory given by (\ref{EMSaction}) by setting $\Psi=0$. The equations of motion look like
\ba
&& \de_\mu(\s{-g}g^{\mu\nu}F_\nu^\rho)=0\ ,\\
&& R_{\mu\nu}-\f{R}{2}g_{\mu\nu}-\f{6}{L^2}g_{\mu\nu}=\f{1}{2}\left(F_{\mu\sigma}F_{\nu}^\sigma-\f{1}{4}g_{\mu\nu}
F_{\rho\lambda}F^{\rho\lambda}\right)\ .
\ea

It is straightforward to obtain the the following solution (see e.g.\cite{HaR})
\ba
&& ds^2=-f(r)dt^2+L^2f(r)^{-1}dr^2+r^2(dx^2+dy^2+dz^2)\ , \\
&& A_t=\mu\left(1-\f{r_+^2}{r^2}\right)\ ,
\ea
with the function $f(r)$ given by
\be
f(r)=r^2\left(1-(b^2+1)\f{r_+^4}{r^4}+b^2\f{r_+^6}{r^6}\right)\ ,
\ee
where we defined $b^2=\f{\mu^2}{3r_+^2}$. The temperature of this black hole
is given by $T=\f{r_+}{2\pi L}(2-b^2)$.

\subsection{Conductivity in Metallic Phase}

It is also interesting to calculate the conductivity in this background. In the AdS$_4$
black hole background, this has been done in \cite{HaR}.
We assume
the fluctuations of $A_x$ and $g_{tx}$ have the time dependence $e^{-i\omega t}$.
We will set $L=r_+=1$ as before.

The Einstein equation for $(x,r)$ component leads to
\be
g'_{tx}-\f{2}{r}g_{tx}-\f{2\mu}{r^3}A_x=0\ , \label{Einxr}
\ee
and the equation of motion for $A_x$ reads
\be
(rf(r)A'_x)'+\f{\omega^2 r}{f(r)}A_x+2\mu
r_+^2\left(\f{g_{tx}}{r^2}\right)'=0\ . \label{MAxx}
\ee

By combining (\ref{Einxr}) and (\ref{MAxx}), we obtain
\be
(rf(r)A'_x)'+\left(-\f{12b^2}{r^5}+\f{\omega^2 r}{f(r)}\right)A_x=0\ .
\ee

\begin{figure}[htbp]
\begin{center}
\includegraphics[height=3.5cm]{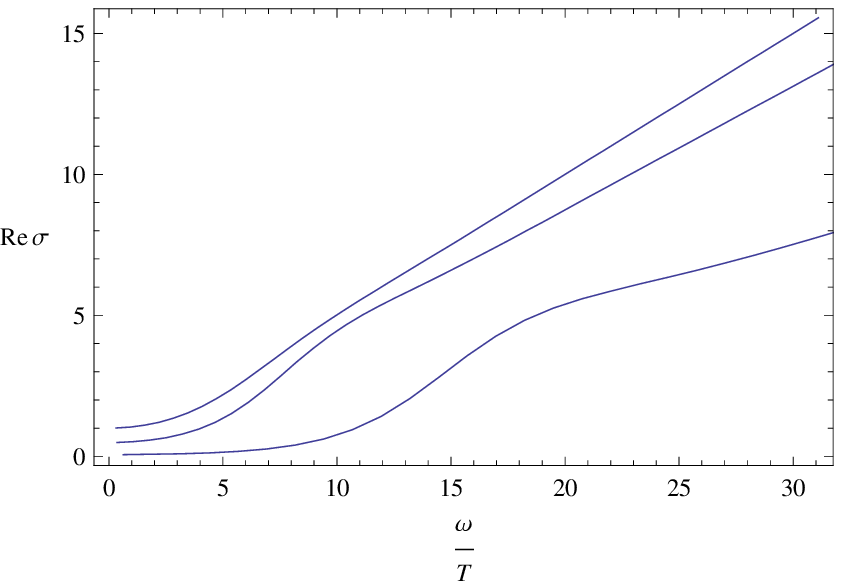}
\hspace{1cm}
\includegraphics[height=3.5cm]{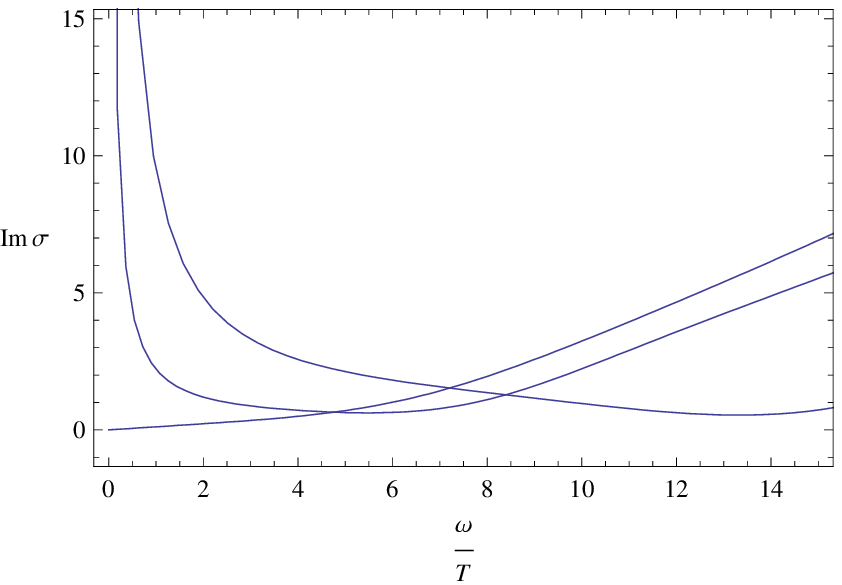}
\end{center}
\caption{The real and imaginary parts of the conductivity in AdS$_5$ charged black hole.
 Three curves correspond to $b=0,\ 0.5$ and $b=1$, respectively in
 the unit $r_+=L=1$. When we fixed $\omega/T$, the real part of the conductivity
decreases as $b$ becomes large, while the imaginary part increases.}
%is raised.}
 \label{fig:AdS5Con}
 \end{figure}

The real and imaginary parts of the conductivity are plotted in Fig.\ \ref{fig:AdS5Con}.
The pole of
Im\,$\sigma(\omega)$ at $\omega=0$ for non-zero values of $\mu$ shows a delta functional contribution to
Re\,$\sigma(\omega)$ via the Kramers-Kronig relation, which should be smoothed by actual
impurities \cite{HHHtwo,HaR}.

\subsection{Holographic Superconductor in AdS$_5$ Black Hole}

In the system in (\ref{EMSaction}), we can write down the equations of motion for
the gauge potential $\Phi$ and the scalar field $\Psi$, which only take real values,  as follows:
\ba
&& \Psi''+\left(\f{f'}{f} + \f{3}{r}
\right)\Psi'+\left(\f{q^2\Phi^2}{f^2}-\f{m^2}{f}\right)\Psi=0\ ,\label{EOMAdS5BH}\\
&& \Phi''+\f{3}{r}\Phi'-\f{2q^2\Psi^2}{f}\Phi=0\ .
\ea

The equation of motion for $A_x$ is given by
\be
A_x''+\left(\f{f'}{f} +
\f{1}{r}\right)A'_x+\left(\f{\omega^2}{f^2}-\f{2q^2\Psi^2}{f}\right)A_x=0\
.
\ee

\begin{figure}[htbp]
 \begin{center}
  \includegraphics[height=3.5cm]{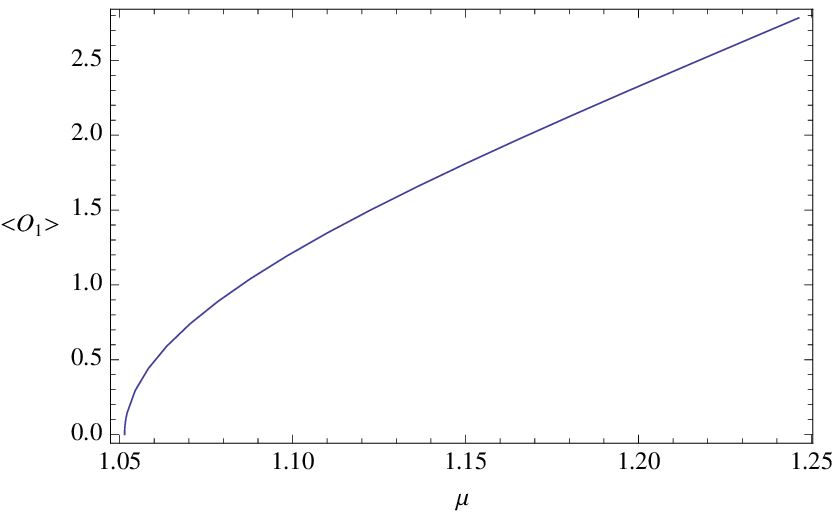}
  \hspace{1cm}
  \includegraphics[height=3.5cm]{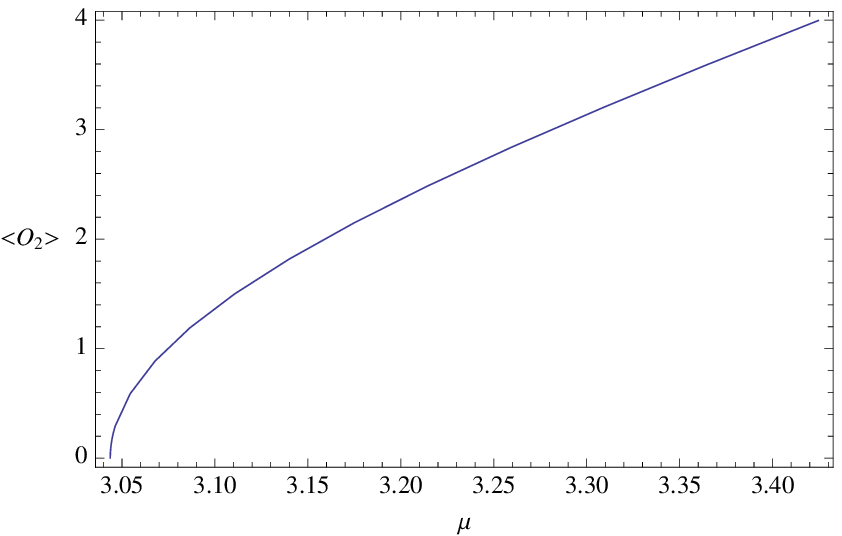}
 \end{center}
 \caption{The scalar condense for $\CO_1$ (left) and $\CO_2$ (right).}
 \label{fig:AdS5BHSC}
\end{figure}

Now we would like to analyze the holographic superconductor phase in this system by numerical calculations.
This system has been studied in \cite{HRone} for different values of $m^2$. Due to the same reason as before, we
set $m^2=-\f{15}{4}$, $L=r_+=q=1$ and $b=0$, which corresponds to the temperature $T=\f{1}{\pi}$.
The result showing the scalar field condensation
is plotted in Fig.\ \ref{fig:AdS5BHSC}.
It shows that the onset of the phase transition into superconductor occurs when
$\mu_1=1.05$ (for $\CO_1$) and $\mu_2=3.04$ (for $\CO_2$).

If we recover $q$ and $T$, then we can conclude that in the probe approximation the
superconductor phase appears when $T<\ap_i q\mu$ due to the condensation of $\la \CO_i\lb$, where
$\ap_1=0.30$ and $\ap_2=0.105$.
The superfluid density in this case behaves like $n_s\sim 40 \m_1
(\m-\m_1)$ for $\CO_1$ and $n_s\sim 2.0 \m_2(\m-\m_2)$ for $\CO_2$,
respectively
\footnote{
For $\CO_2$, the temperature is related with the chemical potential like
$1-\f{T}{T_c} = 1.65 \left( \f{\m}{\m_2} - 1\right)$ around the
critical point $(T=T_c$ and $ \m=\m_2)$. One can obtain $n_s \sim 100 T_c^2
\left( 1 - \f{T}{T_c}\right)$ by using these
relations. This result is consistent with that obtained in \cite{HRone}.}.

\begin{figure}[htbp]
 \begin{center}
  \includegraphics[height=3.5cm]{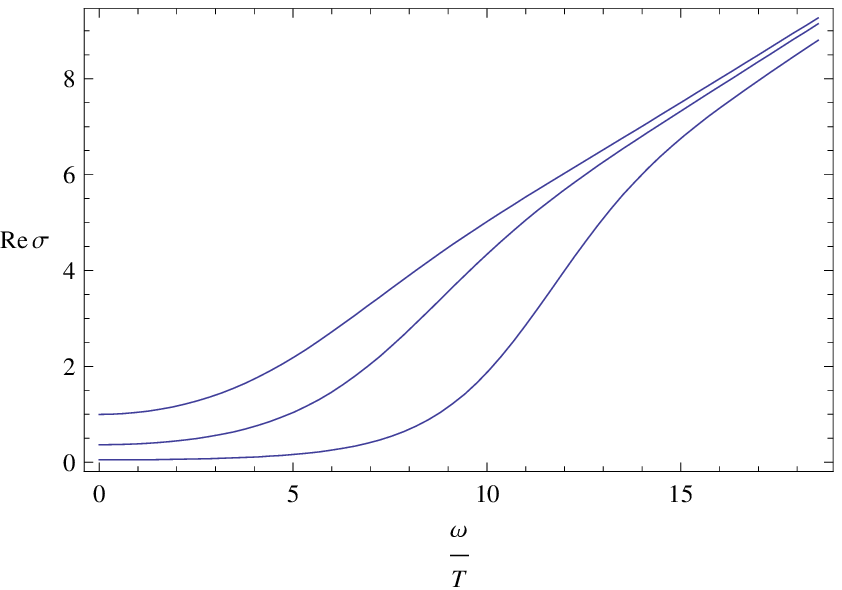}
  \hspace{1cm}
  \includegraphics[height=3.5cm]{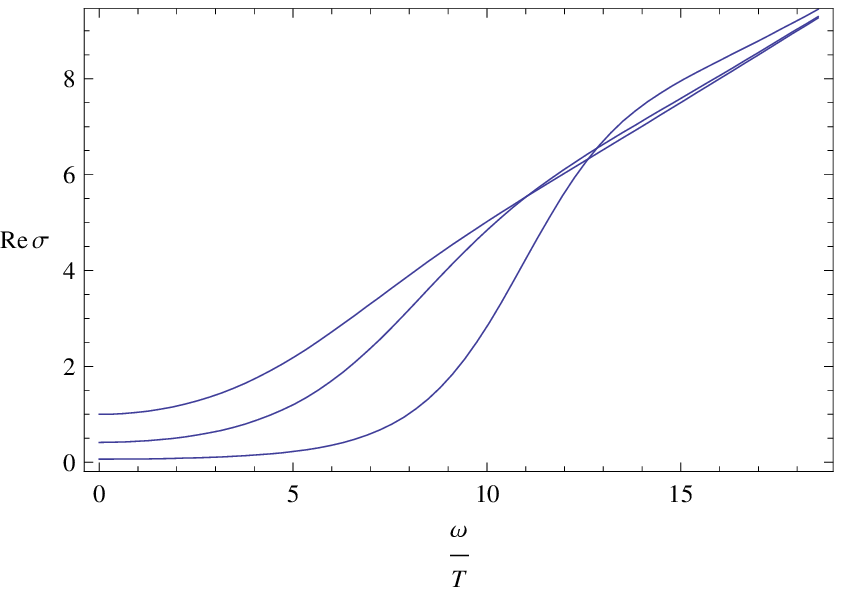}
 \end{center}
 \caption{The real part of the conductivity for $\CO_1\neq 0$ (left) and $\CO_2\neq 0$ (right).
 When we fixed $\omega/T$, the real part of the conductivity
decreases as $|\mu-\mu_c|$ becomes large.
}
 \label{fig:AdS5SCR}
\end{figure}

\begin{figure}[htbp]
 \begin{center}
  \includegraphics[height=3.5cm]{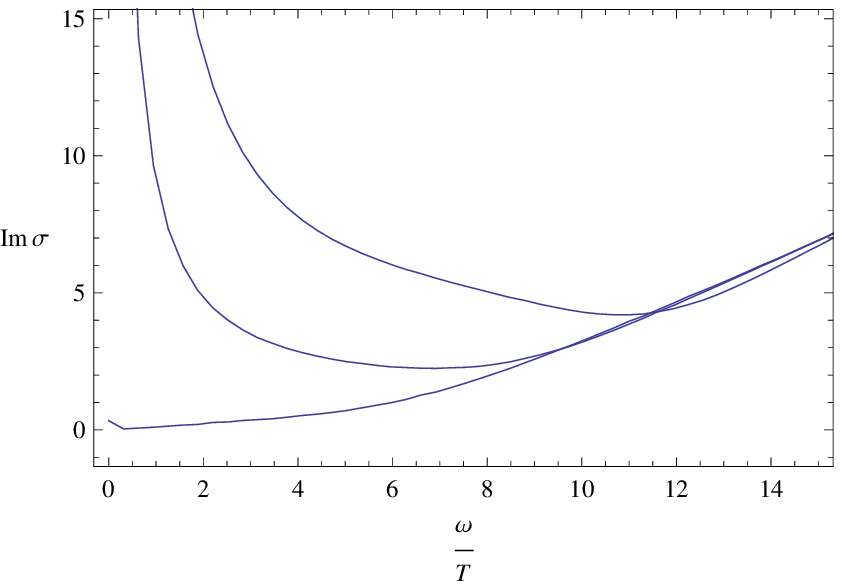}
  \hspace{1cm}
  \includegraphics[height=3.5cm]{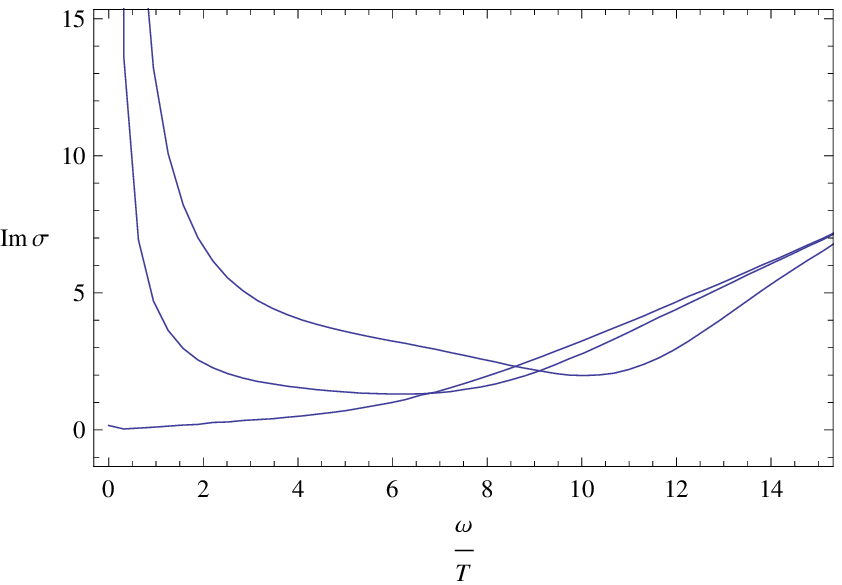}
 \end{center}
 \caption{The imaginary part of the conductivity for $\CO_1\neq 0$ (left) and $\CO_2\neq 0$ (right).
 When we fixed $\omega/T$, the imaginary part of the conductivity
increases as $|\mu-\mu_c|$ becomes large.}
 \label{fig:AdS5SCI}
\end{figure}

The real and imaginary part of conductivity is plotted in Fig.\ \ref{fig:AdS5SCR} and Fig.\ \ref{fig:AdS5SCI}.

\section{Phase Structure}
\label{Phase Structure}

So far, we have studied the superconductor transition both in the AdS$_5$ soliton and
AdS$_5$ BH.\footnote{
We can also consider the pure AdS$_5$ spacetime with the Euclidean time and $\chi$ are compactified. However,
 it has vanishing free energy $\Omega=0$ and is not favored.}
Here we would like to analyze the phase transition between
AdS$_5$ soliton and in AdS$_5$ BH, which is dual to the confinement/deconfinement transition in $2+1$ dimensional
super Yang-Mills \cite{Wi}. After we study this transition,
here we would like to capture the phase diagrams
of our holographic system.

\subsection{Phase transition between AdS$_5$ Soliton and AdS$_5$ Black Hole}

Let us analyze the phase transition between AdS$_5$ soliton and AdS$_5$ charged black hole in the grand canonical
ensemble. The phase boundary is given by the points where the Gibbs Euclidean action $I_G$ of AdS$_5$ soliton coincides with
that of the AdS$_5$ charged black hole. In the gravity description of AdS/CFT, $I_G$ is simply related to
the classical action (\ref{EMSaction}) evaluated for the classical solutions.
Here the scalar field is vanishing. In general it is rewritten as follows
\be
I_G=\beta \Omega=\beta(E-TS-\mu\rho)\ .  \label{grand}
\ee

First consider the AdS$_5$ soliton. In this case,
the only contribution to (\ref{grand}) is the ADM energy $E$ because
$\rho$ and the entropy $S$
are vanishing.
It is evaluated as follows in our normalization\footnote{We set the
Newton constant to $16\pi G_N = 1$ in the action (\ref{EMSaction}).}
\be
\f{\Omega_{sl}}{V_2}=\f{E}{V_2}=-\f{\pi^4 L^3}{R_{0}^3}\ ,\label{osl}
\ee
where $R_0=\f{\pi L}{r_0}$ is the periodicity of $\chi$.  $V_2$ denotes the infinite volume of $(x,y)$ plane.

Next we turn to the AdS$_5$ charged black hole. By employing the result in \cite{HaR}, we find
in our unit
\be
\f{\Omega_{bh}}{V_2}=-\f{r_+^4}{L}\left(1+\f{\mu^2}{3r_+^2}\right)R_0\ . \label{obh}
\ee

We can now analyze the phase transition from (\ref{osl}), (\ref{obh}) and the formula of the temperature
\be
T=\f{r_+}{\pi L}\left(1-\f{\mu^2}{6r_+^2}\right)\ .
\ee
The smaller values of $\Omega$ among  (\ref{osl}) and (\ref{obh}) are favored. It is easy to see that
this phase transition is first order.

\begin{figure}[htbp]
 \begin{center}
  \includegraphics[height=5cm]{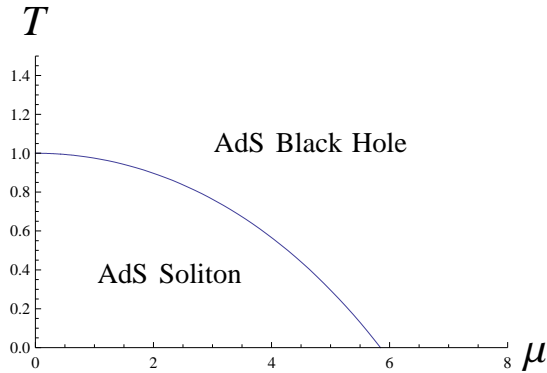}
 \end{center}
 \caption{The phase transition between AdS black hole and AdS soliton. The lower region describes
 the AdS soliton phase, while the other the AdS charged black hole. We
 set $R_0=1$ in this figure.}
 \label{fig:Phasetransition}
\end{figure}

For example, when $\mu$ is vanishing $\mu=0$, the phase transition occurs when $T=\f{1}{R_0}$.
This is obvious since the transition should be the point where the periodicity of Euclidean time
coincides with that of the $\chi$ circle. The AdS black hole and AdS soliton phase correspond to
$TR_0>1$ and $TR_0<1$, respectively. On the other hand, when we increase the value of $\mu$ at zero temperature $T=0$,
a phase transition occurs at
$\mu R_0=2^{\f{1}{2}}3^{\f{1}{4}}\pi\simeq 5.85$ from the AdS soliton to AdS charged black hole. We can also analyze the middle regions
and find the phase diagram in Fig.\, \ref{fig:Phasetransition}.

\subsection{Phase Diagram}

Now we are prepared to write down
our phase diagram
in the probe approximation. We can set $L=1$
without losing generality and also assume $q\gg 1$ so that we can
ignore back reactions to the metric. By the coordinate transformation, we
fix the periodicity of
$\chi$ such as $\chi\sim\chi+\pi$ (corresponding to $r_0=1$ in AdS soliton).
 Thus the independent parameters are $T$ and $\mu$.
The probe approximation is good only when $\mu\ll 1$ as it corresponds to the limit $q\to \infty$ with
$\mu q$ kept finite. Therefore we can reliably write down the phase diagram only
when $\mu\ll 1$. However, notice that the phase boundary between the AdS soliton and its superconductor
is exact even in this approximation and we can
capture important structure of the global phase diagram as we will see below.

\begin{figure}[htbp]
 \begin{center}
  \includegraphics[height=5cm]{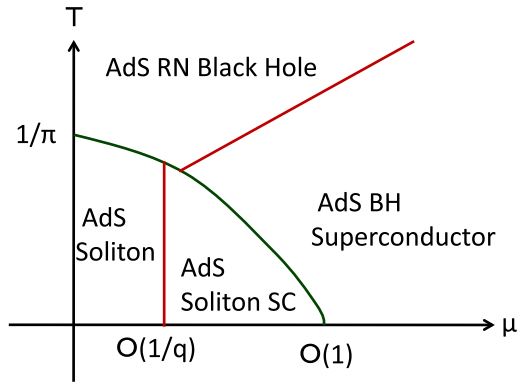}
 \end{center}
 \caption{The phase diagram of AdS soliton and AdS black hole with a charged scalar field obtained in the large $q$ limit.}
 \label{fig: schematic Phase in intro}
\end{figure}

The basic fact is that in the probe approximation, the free energy difference due to the superconductor
phase transition  is of order $O(q^{-2})$ and thus is
much smaller than that of
the transition between the AdS$_5$ soliton and AdS$_5$ BH, which is of order $O(1)$.
In this way, first we may
draw the phase boundary
between the AdS$_5$ soliton and AdS$_5$ BH and then later we can further take into account the superconductor
transition. Finally we obtain the phase diagram as shown in Fig.\, \ref{fig: schematic Phase in intro}.
The relative position of the superconductor curve follows
 from the explicit values $\ap_{1,2}$, $\mu_{1,2}$ and $T_c=\f{1}{\pi}$,
 which separate between various phases as we calculated before by setting $m^2=-15/4$.

\subsection{String Theory Embedding}
In an actual string theory background, we can interpret the charge $q$ as the R-charge of a certain operator in
its dual superconformal field theory. In general, $q$ can be of order one and then the critical chemical potential
$\mu_c$ is also of order one. Thus $\mu_c$ can be greater than the confinement/deconfinement transition point
$\mu_{d}=2^{\f{1}{2}}3^{\f{1}{4}}\simeq 1.86$ assuming $r_0=1$ (or $R_0=\pi$). If this happens, the AdS soliton solution should be
replaced with the AdS black hole at $\mu=\mu_c$ and the superconductor phase transition gets unphysical. To see if we can
avoid this problem in an explicit string theory setup, consider the $AdS_5\times T^{1,1}$ spacetime dual to a
four-dimensional $\mathcal{N}=1$ quiver gauge theory \cite{KW}. The R-charge $R$ with the standard normalization such that
$\Delta=\f{3}{2}R$ for chiral operators is proportional to our charge via the relation $q=\f{\s{3}}{2L}R$ by employing
the analysis of the string theory embedding found in \cite{IIBGH}. Let us concentrate on the chiral operators of the form
Tr$[A_iB_j]$, which have $\Delta=\f{3}{2}$ and $R=1$. For this operator, our previous analysis shows that the superconductor
phase transition in the AdS soliton background occurs at $\mu_c\simeq 0.97$. Notice again that the probe analysis is enough to
fix the transition point even if $q$ is not large enough. Since this value $\mu_c$ is smaller than
$\mu_{d}$, we do not have the problem mentioned before. In this way, we have found an example where we can embed our
zero temperature superconductor/insulator phase transition into string theory.

\subsection{The $t$-$J$ Model and The Slave Boson Approach in RVB Theory}

\begin{figure}[htbp]
 \begin{center}
  \includegraphics[height=5cm]{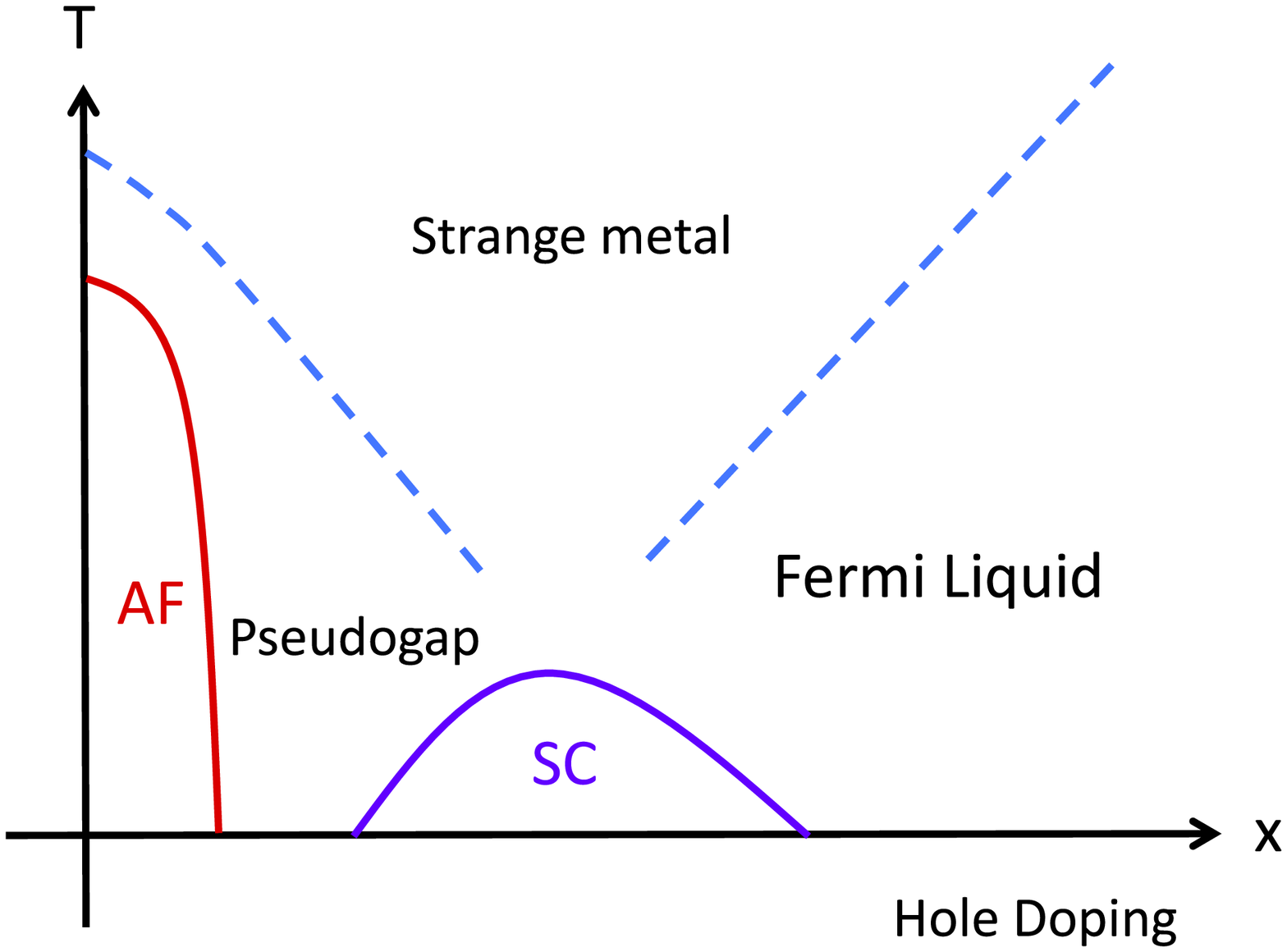}
  \hspace{0cm}
  \includegraphics[height=5cm]{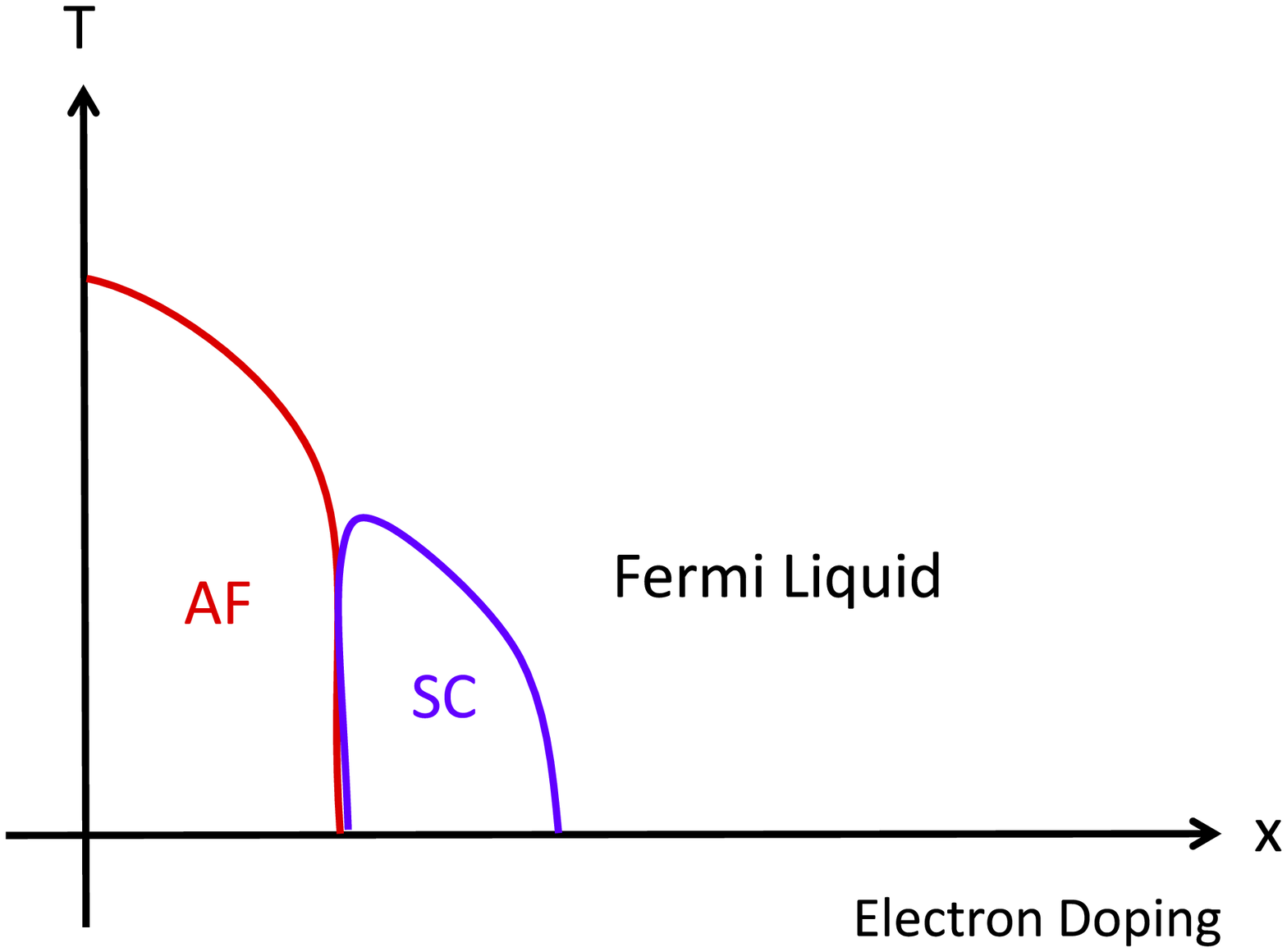}
 \end{center}
 \caption{
The schematic phase diagrams of
the high-$T_c$ cuprate superconductors.
The left panel is for
hole-doped cuprates
(e.g. La$_{2-x}$Sr$_x$CuO$_4$),
while the right is for electron-doped cuprates
(e.g. Nd$_{2-x}$Ce$_{x}$CuO$_4$),
where $x$ measures the amount of doping.
The phase AF denotes the antiferromagnetic phase, which is a Mott insulator;
In the pseudo gap region,
while it is located above the critical temperature,
an energy gap is already observed on the Fermi surface;
The metallic region above the superconducting dome
near the optimal doping is called
the strange metal phase,
which shows anomalous transport properties;
In the Fermi liquid phase
the ground state is adiabatically connected to
the free-fermion ground state with a well-defined Fermi surface.
}
\label{fig:HTC}
\end{figure}

In this section, we compare the superconductor/insulator transition
found in the holographic calculation with the high-$T_c$ cuprates,
in particular with the RVB scenario of high-$T_c$
\cite{LNW}
(see Fig.\  \ref{fig:HTC} for the schematic phase diagram).
At the moment,
there is no single theory
which is capable of
explaining all features of
the high-$T_c$ cuprates
for the entire region of the phase diagram.
Our focus here is on the so-called under-doped region
(the region of phase diagram close to $x=0$),
where the RVB scenario of high-$T_c$ was proposed.
Its applicability to the cuprates, in particular,
beyond the under-doped region,
has been debated.
On the other hand,
for the over-doped region
(the region of phase diagram $x \gtrsim 0.3$),
the Fermi-liquid ground state would be a reasonable starting point.
The electron correlation effects,
in particular,
antiferromagnetic spin fluctuations,
can then be included perturbatively
\cite{Hubbard}.

For an application of the AdS/CFT to high-$T_c$,
see \cite{HaR, nernst,DHS,FLMV},
where in this scenario the anomalous transport in the strange metal phase
is attributed to the quantum critical transport
in the quantum critical regime of the zero temperature quantum critical point.
For other approaches to high-$T_c$, see \cite{ReviewHTC,LNW,Hubbard},
and references therein.
In the following, we would like to suggest a connection between
the AdS/CFT  and the high-$T_c$ superconductors
from a different viewpoint.

The high-$T_c$ cuprate at $x=0$ is the antiferromagnetic insulator.
By doping holes, we frustrate the antiferromagnetic order
and eventually destroy it. Beyond some critical doping $x_c$,
the superconducting ground state emerges.
The physics of
the under-doped cuprates is well-described by the $t$-$J$ model,
\begin{eqnarray}
H =
P \left[
-t \sum_{\langle ij\rangle,\sigma}
c^{\dag}_{i\sigma}
c^{\ }_{j\sigma}
-
\mu \sum_i n_{i}
+
J \sum_{\langle ij\rangle}
\left(
\boldsymbol{S}_i \cdot\boldsymbol{S}_j
-
\frac{1}{4}n_i n_j
\right)
\right]P\ .
\end{eqnarray}
Here, $c^{\dag}_{i\sigma}$ $(c_{i\sigma})$
are an electron creation (annihilation) operator at site $i$ on the square lattice
with spin $\sigma=\uparrow,\downarrow$;
$n_i = \sum_{\sigma} c^{\dag}_{i\sigma} c^{\ }_{i\sigma}$,
and
$\boldsymbol{S}_i = (1/2)
\sum_{\sigma,\sigma'}
c^{\dag}_{i\sigma}\boldsymbol{\sigma}^{\ }_{\sigma\sigma'} c^{\ }_{i\sigma'}$
are the electron number operator,
and the spin operator, respectively;
$t$ represents the kinetic energy for electron (or hole) hopping and $J$ is the antiferromagnetic exchange;
$P$ is the projection operator which removes states with double occupancy at a site $i$
from the physical subspace.
The $t$-$J$ model can be derived from the Hubbard model near half-filling
($x=0$)
by the degenerate perturbation theory, and faithfully captures the physics
near $x=0$.

The major obstacle one encounters when
trying to solve the $t$-$J$ model is the constraint (projection) $P$;
it is nothing but the strong Coulomb repulsion.
The constraint can be analyzed in terms of
the so-called slave-boson approach (gauge theory approach);
it is a calculational tool representing the idea of the RVB,
and for which we make a comparison with our gravity calculations.
In the slave-boson ($SU(2)$ slave-boson) approach,
we decompose electron operators into
bosonic $h_{i}=(b_{i1}, b_{i2})$ and
fermionic $\psi_{i}=(f_{i\uparrow}, f_{i\downarrow}^\dag)$ parts as
\begin{eqnarray}
c^{\ }_{i \uparrow}
\!\!&=&\!\!
\frac{1}{\sqrt{2}}
h^{\dag}_i \psi^{\ }_i
=
\frac{1}{\sqrt{2}}
\left(
b^{\dag}_{1 i} f^{\ }_{\uparrow i}
+
b^{\dag}_{2 i} f^{\dag}_{\downarrow i}
\right)\ ,
\nonumber \\%%%%%
c^{\ }_{i \downarrow}
\!\!&=&\!\!
\frac{1}{\sqrt{2}}
h^{\dag}_i \bar{\psi}^{\ }_i
=
\frac{1}{\sqrt{2}}
\left(
b^{\dag}_{1 i} f^{\ }_{\downarrow i}
-
b^{\dag}_{2 i} f^{\dag}_{\uparrow i}
\right)\ .
\end{eqnarray}
This splitting of electrons
is designed in such a way that it captures the low-energy
degrees of freedom in the problem;
the bosonic field is called holons and roughly describes
the charge degree of freedom of an electron,
whereas the fermionic field is called
spinons and describes the spin degree of freedom of an electron.
When holons and spinons are treated as a free particle,
the Hilbert space for them is larger than the original
Hilbert space, which is three-dimensional per site.
Any physical states $|\mathrm{phys}\rangle$
should belong to the physical
subspace of the fictitious Hilbert space,
and are subjected to the following local constraints:
\begin{eqnarray}
\left(
\psi^{\dag}_i \tau^l \psi^{\ }_i +
h^{\dag}_i \tau^l h^{\ }_i
\right)
|\mathrm{phys}\rangle
=0\ ,
\quad
l=1,2,3,
\quad
\forall i\ .
\end{eqnarray}

The constraint can be implemented in the path integral
by including a Lagrange multiplier $a^l_0$.
The Heisenberg interaction, which takes on the form of a four fermion interaction,
can be decoupled by introducing an auxiliary field $U_{ij}$ defined on a link
connecting site $i$ and $j$.
The partition function in the imaginary time path-integral is then given by
\begin{eqnarray}
Z \!\!&=&\!\!
\int \mathcal{D}\psi \mathcal{D}\psi^{\dag}
\mathcal{D}h^{\dag}
\mathcal{D}h
\mathcal{D}a^1_0
\mathcal{D}a^2_0
\mathcal{D}a^3_0
\mathcal{D}U
\exp\left(- \int^{\beta}_0 d\tau L
\right) \ ,
\end{eqnarray}
where the Lagrangian is given by
\begin{eqnarray}
L
\!\!&=&\!\!
\frac{3J}{8}
\sum_{\langle ij \rangle}
\mathrm{tr}\, \big(
U^{\dag}_{ij} U^{\ }_{ij}
\big)
+
\frac{3J}{8}
\sum_{\langle ij \rangle}
\left(
\psi^{\dag}_i U^{\ }_{ij} \psi^{\ }_j
+
\mathrm{h.c.}
\right)
-\frac{t}{2}
\sum_{\langle ij \rangle}
 \left(
\psi^{\dag}_i h^{\ }_i h^{\dag}_j \psi_j
+
\mathrm{h.c.}
\right)
\nonumber \\%%%%%
&&
+
\sum_i \psi^{\dag}_i
\left(
\partial_{\tau} - {i}
a^{l}_{0 i} \tau^l
\right)\psi^{\ }_i
+
\sum_i h^{\dag}_i
\left(
\partial_{\tau} - {i}
a^{l}_{0 i} \tau^l +\mu
\right)h^{\ }_i \ .
\end{eqnarray}
Integrating over $a^l_0$ gives rise to the constraints.

So far we have just rewritten the $t$-$J$ model.
A typical approach to solve this model, within this representation,
is then to first look for a mean field, and then include
fluctuations around it. The field $a_0^l$ is interpreted
as the time component of a $SU(2)$ gauge field $A^l_\mu$ and its
space components come from the fluctuation of $U_{ij}$.

\begin{figure}[htbp]
 \begin{center}
  \includegraphics[height=5cm]{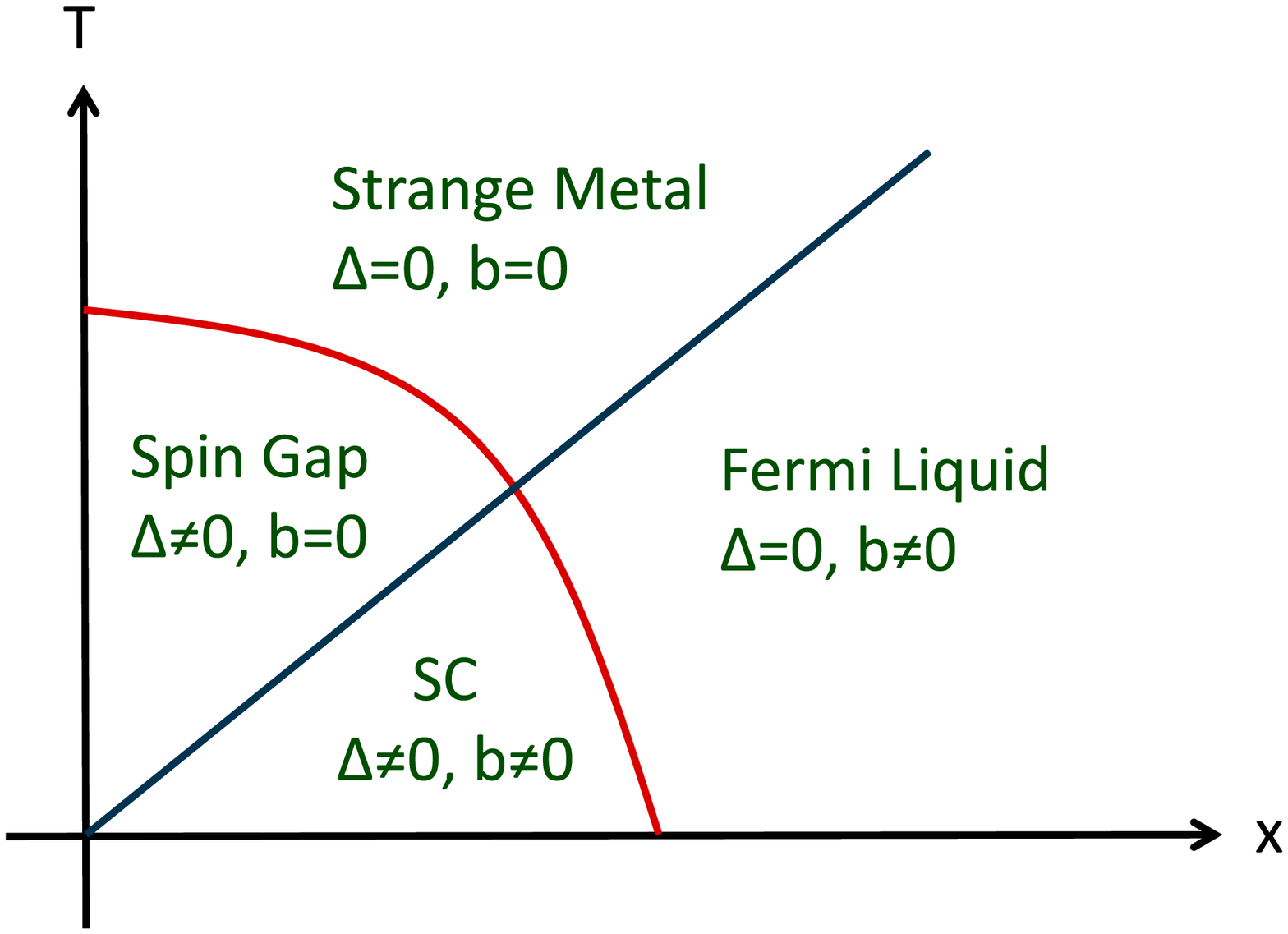}
  \includegraphics[height=5cm]{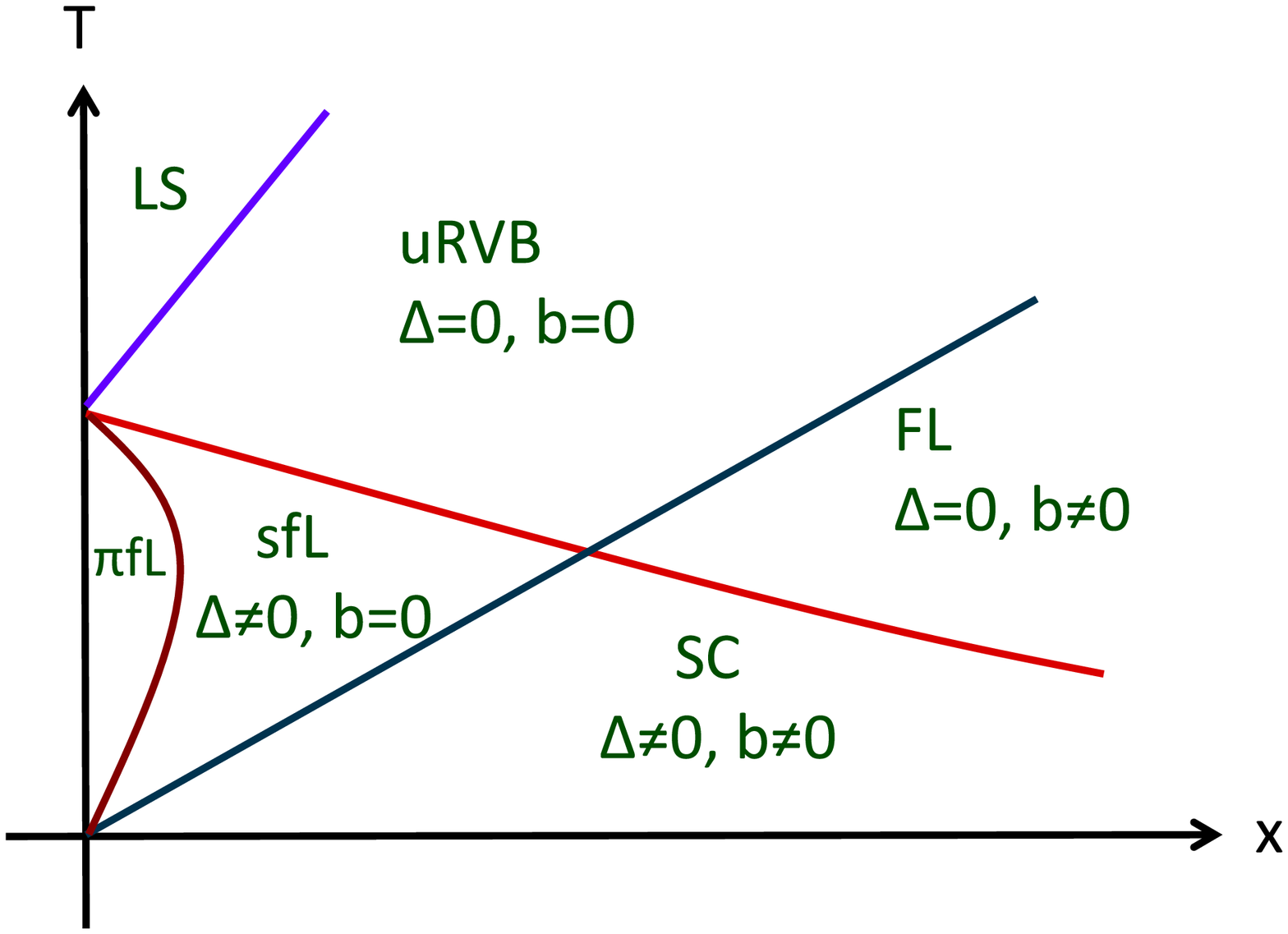}
 \end{center}
 \caption{
The mean field phase diagram as a function of
temperature $T$ and doping $x$ from
the $SU(2)$ (left) and $U(1)$ (right) slave boson theory
\cite{LNW}.
The order parameter $\Delta$ is
 defined by $\Delta_{ij}=f_{i\uparrow}f_{j\downarrow}-f_{i\downarrow}f_{j\uparrow}$
and $b$ represents the expectation value of
the (condensed) holon field.
The $\pi$ flux phase ($\pi$fL) is a spin liquid phase with
the algebraic spin-spin correlation function.
The localized spin (LS) phase is the phase where there are
no hopping matrix elements $U_{ij}=0$ for spinons.
The antiferromagnetic order is
absent in the mean field approximation of the slave boson theory.
}
 \label{fig:RVB}
\end{figure}

The mean field phase diagram from the $SU(2)$ slave boson theory
is shown in Fig.\ \ref{fig:RVB} (left).
(A similar mean field phase diagram can be drawn from the slightly different formalism
of the slave boson theory, the $U(1)$ slave boson theory
[Fig.\ \ref{fig:RVB} (right)]).
By choosing a meanfield (saddle point) configuration for $U_{ij}$,
the low-energy gauge field fluctuations can be
either $SU(2)$, $U(1)$, and $\mathbb{Z}_2$,
since a meanfield configuration $U_{ij}$ can Higgs out partially
the $SU(2)$ gauge fluctuations.

Note that in the mean field phase diagram of the slave-boson approach,
gauge fluctuations are not included.
In the mean field, spinon and holons are treated as an independent
entity, and are deconfined.
Since the gauge coupling is (infinitely) strong,
and since the monopoles that exist
because of the compact nature of the gauge field
could proliferate
we would expect there are confined phases in the phase diagram
although this issue is not completely settled.

\subsection{Comparison Between the Holographic Calculations
and the Slave Boson Approach}

The appearance of the emergent non-abelian gauge group $SU(2)$ in the RVB theory motivates us to
consider its holographic dual gravity description. Indeed, here we would like to argue that our
gravity system discussed in this paper is analogous to the RVB theory of high-$T_c$ superconductors,
though the large $N$ limit has only a qualitative relevance to our problem\footnote{Refer to e.g. \cite{SaRe} for an analysis
assuming the large $N$ limit with the gauge group $Sp(N)$.}.
Before we proceed to the comparison between them, we need to notice that the quantum field theory dual to
our gravity system does not include massless fermions, while the RVB theory
has fermionic spinons.
This might be taken into account in the gravity side
by introducing another (uncharged) scalar field
which is dual to a fermion bilinear $\Delta$ that condenses for small $T$ and $x$.
One possibility of string theory realization
of this will be to add a D7-brane to our D3-brane system\footnote{A simple way to realize this is to wrap
the D7-brane on $t,x,y,r,\chi$ and $S^3\in S^5$. It is easy to see that the fermion bilinear, which is dual to
a mass deformation, condenses in this mode as the D7-brane
should bend due to the bubble of nothing at $r=r_0$. In this case, the fermion condenses just at the confinement/deconfinement
transition point.}. However, here we would like to leave the further analysis for a future work
and to proceed by roughly identifying the confined phase and fermion condensed phase, as the latter is induced to the strongly coupled
gauge interactions.

We first identify the confined phase (the AdS soliton phase)
as a fully-gapped phase of electrons, such as
the valence bond solid (VBS) phase,
or the stripe phase that exists in
$\mathrm{La}_{2-x} \mathrm{Ba}_{x} \mathrm{CuO}_4$
at $x=1/8$.
In the slave boson theory,  such a phase is a confined phase where
spinons and holons are completely glued together, and all
excitations as well as the ground state are well-described in terms
electrons. It may also be possible to identify it with
 the antiferromagnetic phase in the high-$T_c$ cuprates,
where the low-lying excitation is antiferromagnetic magnons,
which is a bound state of two spinons. This gapless
antiferromagnetic magnon may be interpreted as a Nambu-Goldstone
boson induced by the fermion condensation.

The chemical potential $\mu$ in the gravity theory
plays the similar role as doping $x$ in the cuprates;
the chemical potential $\mu$ destroys the confining phase
(the AdS soliton phase) just like
doping $x$ frustrates the antiferromagnetic order.
The phase emerges as we increase $x$ in the cuprates
is superconducting phase; the corresponding phase
in the gravity theory is
the AdS soliton superconductor phase that emerges
as we increase $\mu$.
At the mean field level, the superconducting phase in
the cuprates is the phase where holons are condensed.
In the gravity dual, this is interpreted as the condensation of the scalar field $\Phi$.
Note that in this identification we are not viewing
the scalar field $\Phi$ as the Cooper pair, but a gauge invariant operator made from
two holons.

It is then tempting to identify the deconfining phase
(the AdS RN black hole phase),
lying in the high temperature region above
the AdS soliton and the AdS soliton superconductor phases,
as the pseudo gap phase and the strange metal phase in
high-$T_c$, which crossover with each other.
Indeed, the AdS RN black hole phase,
in the presence of fermions,
shows the non-Fermi liquid behavior
\cite{SSLee,FLMV,Zaanen}
due to the near
horizon AdS$_2$ geometry \cite{FLMV,HKPT}. In the eariler
literature \cite{soo}, the relevance of AdS charged black hole was pointed out
in order to construct the gravity dual of fermi surfaces.
All of these are supportive for our identification of
the AdS RN black hole phase as the pseudo gap or
the strange metal phase, and also
the AdS soliton superconductor phase as
the superconductor phase of the high-$T_c$ cuprates.

One may notice that while the pairing symmetry of the cuprate high-$T_c$ superconductivity
is known to be $d$-wave, there is nothing comparable to it in our gravity calculations.
However, this is not a serious problem as our gravity calculation corresponds to a
effective field theory limit and cannot directly distinguish between
$s$-wave and $d$-wave as is so
in the Ginzburg-Landau theory.

Finally,
the AdS BH superconductor phase that exists
for the large chemical potential can be considered as
the Fermi liquid phase in the over-doped cuprates.
Even though its gravity dual is a superconductor, the
fermions do not condense in our approximation as it is
situated at the deconfined phase. Notice that the expectation value of
Cooper pair (assuming $\la b_{2i}\lb=0$)
\be
\la c_{i\uparrow}c_{j\downarrow}-c_{i\downarrow}c_{j\uparrow}\lb \simeq
\la\Delta_{ij}\lb \la b^{\dag}_{1i}\lb \la b^{\dag}_{1j}\lb \ ,
\ee
is non-vanishing iff both the fermions and holons condense.
Therefore,
this system from the viewpoint of electrons should not be regarded as a superconductor.
Instead, it is an ordinary fermi liquid
 as the spin and charge are tied strongly due to the holon
condensation.

\begin{figure}[htbp]
 \begin{center}
  \includegraphics[height=6cm]{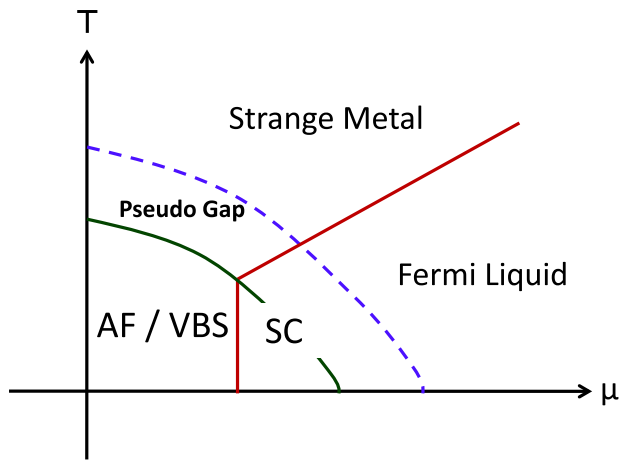}
 \end{center}
 \caption{A speculated phase diagram of our five dimensional Einstein-Maxwell-scalar theory
  with an extra neutral scalar which is dual to a fermion bilinear. The newly added phase boundary which
  signals the fermion pair condensation is written as the dotted blue curve. The other red and green curves denote
    the superconductor and the deconfinement transition, respectively,
as in the same way as in
  Fig.\ \ref{fig: schematic Phase in intro}. }
 \label{fig:finalphase}
\end{figure}

Now let us compare the global phase structure of our holographic system and that of
the RVB theory. At the qualitative level, they look quite similar except that our
system does not actually include the massless fermions (spinons), which makes
the distinction between the pseudo gap phase and the strange metal phase unclear.
Even though we did not realize the fermions, it is not difficult to speculate how the phase structure
of our gravity system changes, for example, by
including an extra scalar field which is uncharged under the
$U(1)$ and is dual to the fermion bilinear. We expect that the fermion pair
condensation continues slightly above the
confinement/deconfinement phase boundary as in the familiar D3-D7 example \cite{KMMMT}.
This is shown in Fig.\ \ref{fig:finalphase} and we filled in the
corresponding phase in high-$T_c$.

One may notice that in hole doped high-$T_c$ superconductors, the antiferromagnetic phase is not
exactly next to the superconductor phase (see Fig.\ \ref{fig:HTC}) and the pseudo gap phase extends
between them as opposed to our holographic system. In this sense, the phase diagram of
electron doped high-$T_c$ superconductors looks more similar to ours.

\section{Conclusions}
\label{Conclusions}

In this paper, we study a five-dimensional
Einstein-Maxwell-scalar field theory with a negative cosmological
constant. We imposed that the boundary of the five-dimensional spacetime approaches
a Scherk-Schwartz compactification of AdS$_5$. The holographic dual should be a certain confining
(2+1)-dimensional gauge theories with $U(1)$ current, which includes the four-dimensional super Yang-Mills
compactified on a circle. We showed that this system undergoes a phase transition
into a superconductor phase (AdS soliton superconductor phase)
due to the charged scalar field condensation when we increase the chemical potential at least for
specific values of the mass and charge of the scalar field. We would like to leave
a more extensive analysis for a future work. We argue that this is a gravity dual of
superconductor/insulator phase transition at zero temperature. We also showed that we can
consistently embed our phase transition into the $\CN=1$ superconformal field theory dual to
AdS$_5\times T^{1,1}$.

We further study the global phase structure of this system by taking into account the
confinement/deconfinement transition at high enough temperature. We compared our phase diagram
(Fig.\ \ref{fig: schematic Phase in intro}) with that of
high-$T_c$ cuprates and we found a qualitative similarity. We also explained
this fact from the viewpoint of the RVB theory of high-$T_c$ superconductors, where
an emergent non-abelian gauge field appears due to the strong Coulomb repulsive force.
This argument suggests that the AdS soliton superconductor phase is dual to
the high-$T_c$ superconductivity phase in the RVB theory.

There are a number of future directions. First of all, since our analysis is restricted to a probe approximation,
it is very interesting to perform a full analysis with backreactions.
 This should make clear the separation between the two superconductor phases.
 Also, to make the comparison with the RVB theory
clear, it is important to take into account massless fermions which play the role of spinons.
In order to obtain more information of
the AdS soliton superconductor, it is useful to exert a magnetic flux in this system. It may also be intriguing to
calculate the entanglement entropy via the AdS/CFT \cite{RT} to understand better the phase transition between
the insulator and superconductor phase because this quantity is non-vanishing even at zero temperature (for the
confinement/deconfinement phase transition, this was done in \cite{NT,KKM}).  Finally, since our results depend on the
values of charges of operators and their conformal dimensions, it will be useful to perform a systematic analysis in concrete
string theory setups.

\vskip6mm
\noindent
{\bf Acknowledgments}

It is a great pleasure to thank T. Eguchi, G. Horowitz, H. Liu,
Y. Ran for useful discussions. We are very grateful to Y. Hikida and S.-S. Lee
for comments on the draft of this paper. TN and TT is supported
by World Premier International Research Center Initiative
(WPI Initiative), MEXT, Japan.
The work of TN is supported by JSPS Grant-in-Aid for Scientific Research No.\,19$\cdot$3589.
The work of TT is also supported in
part by JSPS Grant-in-Aid for Scientific Research No.20740132, and
by JSPS Grant-in-Aid for Creative Scientific Research No.\,19GS0219.
SR thanks the Center for Condensed Matter Theory at University of California,
Berkeley for its support.

\noindent
%{\bf Note added}:

\vskip2mm

\appendix

\section{Schr{\" o}dinger Potential Description}
\label{sc:ScPot}

\subsection{AdS Soliton}

\begin{figure}[htbp]
\begin{center}
\includegraphics[height=4cm]{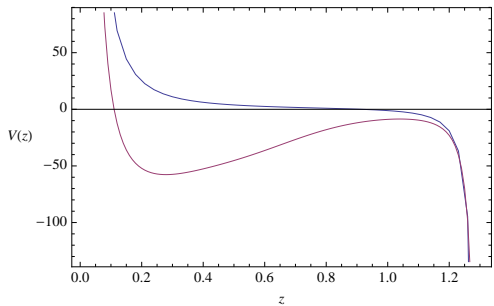}
\end{center}
\caption{The potential $V(z)$ for the AdS soliton without (blue) and
 with (red) the
 scalar operator condensation.}
\label{fig:Pot}
\end{figure}

The condensation of the scalar operator can be interpreted as the instability of the scalar
field around the tip of the AdS soliton solution as follows.
The equation of motion of the scalar field $\Psi$ with a time dependence
$e^{-i\omega t}$ gives
\begin{align}\label{ScInst}
 \Psi'' + \left( \f{f'}{f} + \f{3}{r}\right)\Psi' + \left( -\f{m^2}{f}
 + \f{\Phi^2 + \omega^2}{r^2f}\right)\Psi = 0 \ .
\end{align}
This equation can be transformed to the Schr{\" o}dinger equation by
introducing a new radial coordinate
\begin{align}
 z&= \int_r^{\infty}\f{ds}{s\s{f(s)}} = \f{{}_2F_1(\f{1}{4},\f{1}{2},\f{5}{4},\f{1}{r^4})}{r}\ ,
\end{align}
and redefining the scalar field like $\Psi\equiv B\psi$ with
appropriate choice of the function $B$, which is given by $B(z)=r(z)^{-1/2}|r'(z)|^{-1/2}$.
The equation becomes
\begin{align}\label{SchEq}
 -\p_z^2\psi(z) + V(z)\psi = \omega^2 \psi(z) \ ,
 \end{align}
where the potential is calculated to be
\be
V(z) =\f{(15+4m^2)r(z)^8-2(9+2m^2)r(z)^4-1}{4r(z)^2(r(z)^4-1)}-q^2\Phi^2
\ . \label{SPot}
\ee
This potential $V(z)$ is plotted in Fig.\ \ref{fig:Pot} with and witout
the scalar operator condensation, i.e., $\Phi$ takes zero and nontrivial
values, respectively. The range of $z$ is $0<z\leq z_*$, where
$z_*=\f{\s{\pi}\Gamma(5/4)}{\Gamma(3/4)}(\simeq 1.31)$.

We can find the following behaviors:
\ba
&& V(z)\simeq \f{m^2+\f{15}{4}}{z^2} \ \ \ \ (z\to 0) \ ,\label{BFZ} \\
&& V(z)\ \ \simeq -\f{1}{4(z-z_*)^2} \ \ \ \ (z\to z_*) \ . \label{TIPI}
\ea

It is well-known that the Schr{\" o}dinger problem for the potential
$V(z)=\f{k}{z^2}$ is ill-defined (or unstable) for $k<-1/4$ as this leads to infinitely many
negative energy states. When applied to (\ref{BFZ}), this leads to the BF bound $m^2\leq -4$.
On the other hand, the behavior near the tip
(\ref{TIPI}) shows that the system is marginally stable. In addition, the actual potential is
larger than the right-hand side of (\ref{TIPI}) with the high potential wall near $z=0$.
Thus the ground state energy $\omega^2$ takes a finite positive value and this explains the
mass gap in the AdS soliton. The introduction of the gauge potential $\Phi$ decreases the ground
state energy and above a certain value, the system becomes unstable under the scalar condensation
in the IR region.

\subsection{AdS Black Hole}

\begin{figure}[h]
\begin{center}
\includegraphics[height=4cm]{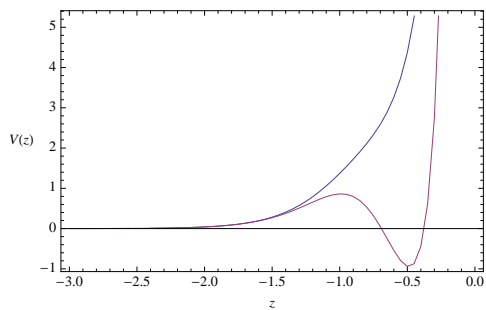}
\end{center}
\caption{The potentials $V(z)$ for the AdS BH without (blue) and with
 (red) the scalar
 operator condensation.}
\label{fig:PotAdSBH}

\end{figure}

It is also useful to analyze the AdS black hole background in a similar way. We will set $\mu=0$ for
simplicity. Introducing the new radial coordinate $z$ ($-\infty<z<0$)
\begin{align}
 z&= -\int_r^\infty \f{ds}{f(s)} = \f{1}{4}(-\pi - 2\coth^{-1} (r) +
 2\tanh^{-1}(r)) \ ,
\end{align}
the equation of motion for the scalar field (\ref{EOMAdS5BH}) is simplified to the Schr{\" o}dinger
equation (\ref{SchEq}) with
$B(z)=r(z)^{-3/2}$.
The potential becomes
\begin{align}
 V(z)= \f{(r(z)^4-1)\left((15+4m^2)r(z)^4+9\right)}{4r(z)^6}-q^2\Phi^2 \
 .
\end{align}
This always vanishes on the horizon ($f=0$) when $\Phi=0$
with arbitrary choice of $m^2$.
The form of this potential is depicted in Fig.\, \ref{fig:PotAdSBH} with
and without the scalar operator condensation.

\vspace{1cm}

%\newpage

%%%%%%%%%% References %%%%%%%%%%%%%%%%%%%%%%%%%
\newcommand{\J}[4]{{\sl #1} {\bf #2} (#3) #4}
\newcommand{\andJ}[3]{{\bf #1} (#2) #3}
\newcommand{\AP}{Ann.\ Phys.\ (N.Y.)}
\newcommand{\MPL}{Mod.\ Phys.\ Lett.}
\newcommand{\NP}{Nucl.\ Phys.}
\newcommand{\PL}{Phys.\ Lett.}
\newcommand{\PR}{ Phys.\ Rev.}
\newcommand{\PRL}{Phys.\ Rev.\ Lett.}
\newcommand{\PTP}{Prog.\ Theor.\ Phys.}
\newcommand{\hep}[1]{{\tt hep-th/{#1}}}
%%%%%%%%%%%%%%%%%%%%%%%%%%%%%%%%%%%%%%%%%%%%%%%

\end{document}